\documentclass[traditabstract]{aa} 
\usepackage{graphicx}
\usepackage{txfonts}
\begin{document}

\title{Morphology with Light Profile Fitting of Confirmed Cluster Galaxies at z=0.84}
\author{Julie B. Nantais\inst{1}, Hector Flores\inst{2}, Ricardo Demarco\inst{1}, Chris Lidman\inst{3}, Piero Rosati\inst{4}, \& M. James Jee\inst{5}}

\institute{Departamento de Astronom\'ia, Universidad de Concepci\'on, Casilla 160-C, Concepcion, Chile
\and
GEPI, Paris Observatory, CNRS, University of Paris-Diderot; 5 Place Jules Janssen, 92195 Meudon, France
\and
Australian Astronomical Observatory, PO Box 296, Epping NSW 1710, Australia
\and
ESO, Karl-Schwarzchild Strasse 2, 85748, Garching, Germany
\and
Department of Physics, University of California, Davis, One Shields Avenue, Davis, CA 95616, USA
}

   \date{Received ???; accepted ???}

\abstract{We perform a morphological study of 124 spectroscopically confirmed cluster galaxies in the z=0.84 galaxy cluster RX J0152.7-1357.  Our classification scheme includes color information, visual morphology, and 1-component and 2-component light profile fitting derived from Hubble Space Telescope $riz$ imaging.  We adopt a modified version of a detailed classification scheme previously used in studies of field galaxies and found to be correlated with kinematic features of those galaxies.  We compare our cluster galaxy morphologies to those of field galaxies at similar redshift.  We also compare galaxy morphologies in regions of the cluster with different dark-matter density as determined by weak-lensing maps.  We find an early-type fraction for the cluster population as a whole of 47\%, about 2.8 times higher than the field, and similar to the dynamically young cluster MS 1054 at similar redshift.  We find the most drastic change in morphology distribution between the low and intermediate dark matter density regions within the cluster, with the early type fraction doubling and the peculiar fraction dropping by nearly half.  The peculiar fraction drops more drastically than the spiral fraction going from the outskirts to the intermediate-density regions.  This suggests that many galaxies falling into clusters at z$\sim$0.8 may evolve directly from peculiar, merging, and compact systems into early-type galaxies, without having the chance to first evolve into a regular spiral galaxy.}


  \titlerunning{Morphology of cluster galaxies at z = 0.84}
  \authorrunning{Nantais et al.}

  \maketitle

\section{Introduction}

For about as long as galaxies outside the Milky Way have been recognized as distinct entities, galaxies in the local universe have been classified according to the Hubble sequence (Hubble \cite{hub26}) as elliptical, lenticular, spiral, and irregular.  The former two categories are called ``early types'' and are generally low in star formation.  Spirals are often called ``late types'' and typically have moderate to high star formation rates.  Irregular galaxies, though having a variety of morphologies and origins and being off the main Hubble sequence, may have high star formation rates as well.  Dressler (\cite{dre80}) and others found that there is a consistent relationship in the local universe between local galaxy density and the frequency of these different galaxy morphologies.  Elliptical and lenticular galaxies are most common in high-density regions such as the cores of galaxy clusters, while spiral and irregular galaxies are more predominant in small groups and the field.  Studies at higher redshift indicate that this morphology-density relation has been in place for over half the age of the Universe.  It is commonly found up to z $\sim$ 1 in both mass-selected samples (Holden et al.~\cite{hol07}; van der Wel et al.~\cite{van07}), and luminosity selected samples (Postman et al.~\cite{pos05}; Smith et al.~\cite{smi05}) and even up to z $\sim$ 1.46 (Hilton et al.~\cite{hil09}).  In the mass-selected samples (Holden et al.~\cite{hol07}; van der Wel et al.~\cite{van07}), no evolution in the early-type fraction is found among massive and intermediate-mass cluster galaxies since z = 0.8.  Field studies suggest that in these environments as well, early-type galaxies formed at high redshift (Cameron et al.~\cite{cam11}) and their numbers have remained largely stable since intermediate redshift (Delgado-Serrano et al.~\cite{del10}).

Although it is consistently found that early-type galaxies are more predominant in clusters than in the field, the actual ratio of early-type galaxies to late-type and irregular galaxies depends on the method of classification and choice of sample.  In Holden et al.~(\cite{hol07}) and van der Wel et al.~(\cite{van07}), early type fractions of visually classified, mass-selected samples were around 80-90\% in clusters and about 40\% in the field.  However, van Dokkum et al.~(\cite{vando00}) found only 44\%  undisturbed early-type galaxies in the z $\sim$ 0.83 galaxy cluster MS 1054 with a classification system that included probable merging systems.  Similarly, Delgado-Serrano et al.~(\cite{del10}) found only 17\%  of field galaxies were early types at z $\sim$ 0.6 when accounting for mergers, tidal disruption, and blue nuclei.

The best method of studying the physical process of galaxy evolution is to be able to map a galaxy's kinematics.  This has been done successfully with optical and near-infrared integral field spectroscopy for star-forming field galaxies at redshifts between 0.4 and 0.8 (Flores et al.~\cite{flo06}; Puech et al.~\cite{pue06a}; Puech et al.~\cite{pue06b}; Puech et al.~\cite{pue08}; Yang et al.~\cite{yan08}; Hammer et al.~\cite{ham09}; Puech et al.~\cite{pue09}) and between redshifts 0.9 and 3 (Bouch\'e et al.~\cite{bou07}; Genzel et al.~\cite{gen08}; Epinat et al.~\cite{epi09}; F\"orster Schreiber et al.~\cite{for09}; Law et al.~\cite{law09}; Wright et al.~\cite{wri09}; Mancini et al.~\cite{man11}; Alaghband-Zadeh et al.~\cite{ala12}; Contini et al.~\cite{con12}; L\'opez-Sanjuan et al.~\cite{lop12}; Queyrel et al.~\cite{que12}; Vergani et al.~\cite{ver12}).  These studies demonstrated that field galaxies at intermediate to high redshift typically have somewhat perturbed kinematics and thus have not yet settled into the regular spiral disk systems commonly seen in the field today.   Kinematic studies can also be performed with slit spectroscopy of star-forming galaxies if the galaxy is sufficiently spatially resolved along the axis observed in the slit.  Jaff\'e et al.~(\cite{jaf11}) found evidence for kinematic disturbance among morphologically regular disk galaxies falling into clusters up to z = 1 with well-resolved slit spectroscopy.  However, for galaxies with low star formation rates, integral field spectroscopy can be difficult to perform outside the low-redshift universe.  The galaxy's continuum and faint emission or absorption lines may be too faint to obtain adequate signal-to-noise in the small spaxels of an integral field unit.

The next best way to study detailed galaxy morphology would be to identify morphological features in images that are known to be strongly correlated to genuine kinematic differences among galaxies.  The Neichel et al.~(\cite{nei08}) study used multiple photometric bands plus kinematic information to create an enhanced morphological classification scheme.  Their main objective was to differentiate regular spiral galaxies from various kinds of peculiar galaxies and mergers, and correlate the morphological peculiarities with kinematic ones.  They found notable differences in the kinematics of their regular spirals and their various peculiar categories.  Delgado-Serrano et al.~(\cite{del10}) expanded on this classification system to include early-type galaxies (elliptical and S0) and compare the Hubble sequence of field galaxies at z$\sim$0.6 with morphologies of z=0 galaxies.  Both samples were mass-selected and primarily intermediate-mass, as opposed to many high-redshift samples which are magnitude-selected in the rest-frame ultraviolet.  They found that over the course of 6 Gyr, peculiar, compact, and merging systems must have systematically evolved into regular spiral galaxies to create the distribution of morphologies we see today.  Hammer et al.~(\cite{ham05}) used multi-wavelength photometry, including infrared, to determine the star formation and merger history of galaxies at intermediate redshifts (z$>$0.4) and came to a similar conclusion.  According to Hammer et al.~(\cite{ham05}), most field disk galaxies had undergone mergers since z = 1, and many of these galaxies passed through a Luminous Infrared Galaxy (LIRG) phase.

The morphological studies of Neichel et al.~(\cite{nei08}) and Delgado-Serrano et al.~(\cite{del10}), as well as studies performed thus far with integral field spectroscopy, have focused on galaxy evolution in field environments.  However, as the Morphology-Density relation shows, field galaxies have a different history of galaxy evolution from those in clusters.  Studying the morphological evolution of galaxies in and around rich clusters, in a manner likely to reflect the galaxy's kinematics, is necessary to understand environmental effects on galaxy evolution.  RX J0152.7-1357, hereafter J0152.7, is a dynamically young (Demarco et al.~\cite{dem05}; Girardi et al.~\cite{gir05}), intermediate-redshift cluster (z $=$ 0.84).  With its 134 spectroscopically confirmed cluster members (Demarco et al.~\cite{dem05}, \cite{dem10}) of all morphologies, it is an excellent place to look for the evolution of star-forming galaxies into passive early types.  In color-magnitude terms, this is typically expressed in terms of evolution from the mostly spiral and irregular ``blue cloud'' to the mostly early type ``red sequence.''  Demarco et al.~(\cite{dem10}) have found evidence that some of the less massive red-sequence galaxies in the cluster have only recently evolved onto the red sequence.  Patel et al.~(\cite{pat09}) also found evidence for recent and ongoing evolution onto the red sequence in spectroscopically confirmed galaxies with $i' < 23.75$ in infall groups.

Several morphological studies on J0152.7 (Postman et al.~\cite{pos05}; Blakeslee et al.~\cite{bla06}; Holden et al.~\cite{hol07}) have previously been performed, but only with an interest in distinguishing early-type from late-type visual morphologies in a general sense.  Their classification schemes did not attempt to account for regular vs. perturbed galaxy kinematics like the field galaxy classification scheme in Neichel et al.~(\cite{nei08}) or Delgado-Serrano et al.~(\cite{del10}).  Thus, a study of the morphology of galaxies in J0152.7 using a classification scheme similar to that of Delgado-Serrano et al.~(\cite{del10}) could provide a richer picture of the state of galaxy evolution and its relation to environment in this cluster.

The main goal of this paper is to improve our understanding of how the cluster environment at intermediate redshift affects the state of galaxy evolution.  We intend to classify galaxies not only as early types or late types, but also by whether they have AGN, nuclear starbursts, tidal distortions, or mergers.  We will compare our style of morphology analysis to previous work in which the latter features were not considered.  In Section 2 of this paper, we describe the data used in our analysis.  In Section 3 we describe the methodology used to classify our galaxies.  In Section 4 we compare the morphologies of subsamples of our cluster galaxies as a function of environment and stellar mass.  We also compare these results to field galaxy studies and other cluster galaxy studies at the same redshift, and discuss the implications of our results for galaxy evolution.  And finally, we provide our conclusions in Section 5.  Throughout this paper, we assume a $\Lambda$CDM cosmology with $\Omega_M$ $=$ 0.3, $\Omega_{\lambda}$ $=$ 0.7, and $H_0$ $=$ 70 km s$^{-1}$ Mpc$^{-1}$.

\section{Data and Object Selection}
We used the Hubble Space Telescope/Advanced Camera for Surveys, hereafter HST/ACS (Ford et al.~\cite{for98}) imaging of J0152.7 associated with HST proposal 9290 (PI: Holland Ford).  These images were taken in the broadband filters F625W ($r_{625}$), F775W ($i_{775}$), and F850LP ($z_{850}$).  The images were drizzled as described in Blakeslee et al.~(\cite{bla06}), and have a pixel scale of approximately 0.05 arcsec pixel$^{-1}$.  The details of the ACS observations and how they were reduced are explained in detail in Blakeslee et al.~(\cite{bla06}) and described more briefly in Table 1 of Postman et al.~(\cite{pos05}) and in Demarco et al.~(\cite{dem10}).  The area of the $riz$ mosaics was about 36.5 square arcmin (Postman et al.~\cite{pos05}) centered on the core of the northern sub-cluster, and the exposure time was approximately 4800 s per frame.  Since the frames overlapped, creating about a 1 arcmin overlap region in the center, some areas had 2 to 4 times the typical exposure time per frame.  The surface brightness values corresponding to 1 $\sigma$ above the sky background value were 25.33 AB mag arcsec$^{-2}$ in $r_{625}$, 25.15 AB mag arcsec$^{-2}$ in $i_{775}$, and 24.84 AB mag arcsec$^{-2}$ in $z_{850}$.  These surface brightness limits were calculated using the mean and $\sigma$ of a region in the eastern part of each mosaiced image that was free of objects in all three bands.

The bulk of the 134 confirmed members of J0152.7 were identified by Demarco et al.~(\cite{dem05}, \cite{dem10}) using VLT FORS1 and FORS2 spectroscopy.  The original spectroscopic targets were chosen to have $R$ $<$ 24 and 0.7 $<$ z$_{phot}$ $<$ 0.95.  We found 124 of these galaxies within the area covered by the HST/ACS images, covering a wide range of stellar masses (Holden et al.~\cite{hol07}) and morphological T-types (Postman et al.~\cite{pos05}).  Demarco et al.~(\cite{dem10}) have determined $r_{625}$, $i_{775}$, and $z_{850}$ magnitudes for each of these 124 galaxies.  The faintest confirmed cluster member in the $z_{850}$ band has a magnitude of 25.56, though all but two have $z_{850}$ $<$ 24.

\section{Light Profile Fitting, Color Maps, and Galaxy Classification}
\subsection{Galaxy Light Profile Fitting}
We used the GALFIT software (Peng et al.~\cite{pen02}) to fit a S\'ersic bulge and an exponential disk (S\'ersic index fixed at 1) component to each galaxy.  Various input parameters for the GALFIT light profiles were computed using Source Extractor (Bertin \& Arnouts \cite{ber96}) version 2.8.6.  We basically followed the methodology of Neichel et al.~(\cite{nei08}) and Delgado-Serrano et al.~(\cite{del10}), masking neighboring objects to reduce flux contamination.  We used Source Extractor to derive initial guesses for the central coordinates, size, sky background, and other input parameters, and fit a 301 $\times$ 301 pixel (15 $\times$ 15 arcsec) area around each galaxy.  The masks for the neighboring objects were created by inverting the Source Extractor segmentation maps (so that objects would be masked and the sky would be revealed) and then removing the portion corresponding to the object being fit.  Like Delgado-Serrano et al.~did for their higher-redshift galaxy sample, we fit and analyzed the galaxies in the $z_{850}$-band.  We identified 23 bright to intermediate magnitude, non-saturated, relatively isolated stars throughout the mosaic to create an empirical point spread function (PSF) using DAOPHOT in IRAF$^{1}$.  Since the PSF in the $z_{850}$-band had a full-width at half-maximum (FWHM) of about 1.7 pixels, we double-sampled our PSF with SEEPSF in IRAF to achieve Nyquist sampling (FWHM $\geq$ 2 pixels).

\footnotetext[1]{IRAF is distributed by the National Optical Astronomy Observatory, which is operated by the Association of Universities for Research in Astronomy, Inc., under cooperative agreement with the National Science Foundation.}

We have made several changes to the methodology of light profile fitting to deal with the difficulties of obtaining a good fit in a crowded field.  First, rather than uniformly fixing the background sky value at the value derived by Source Extractor, we allowed it to remain a free parameter when possible.  This is because the background can be difficult to calculate and can have a noticeable effect on the stability and quality of a fit in a crowded field.  Secondly, for the reasons mentioned above, we used a double-sampled PSF rather than a single-sampled one.  We specified in our GALFIT parameter files that the PSF was double-sampled relative to the data.  Thirdly, whenever doing so yielded a stable fit, we constrained the S\'ersic index of the bulge to 1.5 $<$ n $<$ 6.0 to avoid large errors in the bulge half-light radius and magnitude.  Finally, we created a hierarchy of fits with certain parameters fixed that we would use whenever the default fit proved unstable.  This proved necessary because, for 61 out of 124 galaxies, a fit with the default set of free parameters had either multiple GALFIT parameters out of range or the center of the bulge or disk located well outside the galaxy. 

\begin{figure}
\centering
\includegraphics[width=4.25cm]{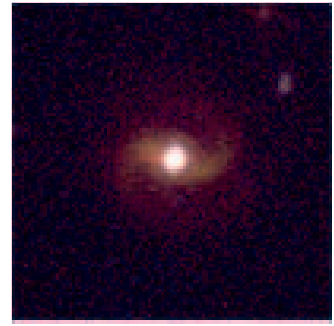}
\caption{5 arcsec by 5 arcsec image of ID 300, a galaxy with a bright active nucleus (Demarco et al.~\cite{dem05}).  It is the only galaxy for which a one-component light profile fit failed to converge.}
\label{FigGal300}
\end{figure}

\begin{table*}
\caption{\label{t1}Variation of GALFIT Fitting Parameters}
\centering
\begin{tabular}{lccc}
\hline\hline
Fit Number & Bulge constraints & Position constraints & Sky constraints \\ 
\hline

1 & 1.5$<$n$<$6 & none & none \\
2 & 1.5$<$n$<$6 & fixed at Source Extractor value & none \\
3 & 1.5$<$n$<$6 & none & fixed at Source Extractor value \\
4 & n$=$4 & none & none \\
5 & n$=$4 & fixed at Source Extractor value & none \\
6 & n$=$4 & none & fixed at Source Extractor value \\
7 & none & none & none \\
8 & none & fixed at Source Extractor value & none \\
9 & none & none & fixed at Source Extractor value \\
\hline
\end{tabular}

\end{table*}

\begin{figure*}
\centering
\includegraphics[width=18cm]{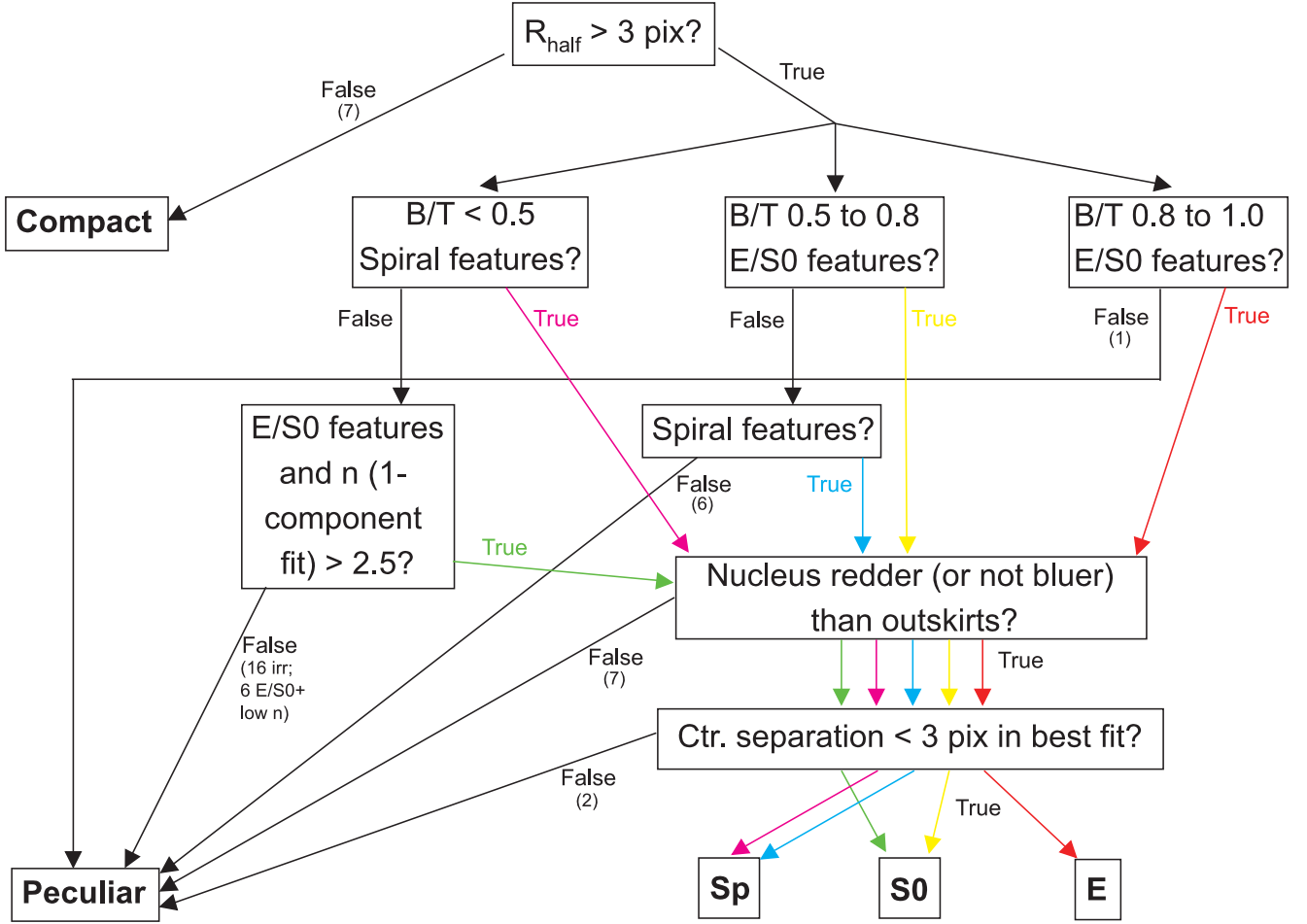}
\caption{New decision tree for morphology classification.  Classification system is modified from earlier versions (Neichel et al.~\cite{nei08}, Delgado-Serrano et al.~\cite{del10}) to account for the consequences of poor 2-D fits in making the S0 vs. spiral distinction.  Numbers of galaxies classified as Peculiar or Compact by means of a particular criterion are shown with the final ``False'' arrows that eliminate the galaxy from regular E/S0/Spiral classification.}
\label{FigDecTree}
\end{figure*}

The hierarchy of parameter constraints used to perform two-component light profile fitting is shown in Table 1.  Ideally, a final fit gave the bulge and disk centers within about 3 pixels of each other, no parameters out of range in GALFIT, and no half-light radii exceeding 150 pixels.  Fit 1 was tried before all other fits, and Fit 9 was tried if all other fits ``failed''  - i.e., did not meet the above criteria for a successful fit.  If none of the fits met the above criteria, the one with the fewest of the above numerical problems was chosen as the final fit.  We prioritized fixing the center over fixing the sky background because we expected most bulge and disk centers to be fairly well-aligned, and because of the difficulty of calculating the sky in crowded fields.  We prioritized fixing the centers and the sky over fixing the S\'ersic index because we did not necessarily expect all bulges to follow a perfect de Vaucouleurs law.

Ultimately, only 63 galaxies (just over half of the total sample) could be fit with two components using the default Fit 1.  Forty galaxies - nearly a third - could only be fit with Fit 3 or higher, or had numerical problems that could not be resolved by any fit.  Because of this, along with a surprisingly high number of galaxies (about 24) with regular early-type visual features having bulge-to-total ratios (B/T) of less than 0.5, we also performed one-component fits.  For the one-component fits, we used GALFIT in the style of Blakeslee et al.~(\cite{bla06}) but with the sky left to vary when possible.  We used Blakeslee et al.'s constraints on the S\'ersic index, 1 $<$ n $<$ 4, and also, as they did, masked out a region with a 2-pixel radius in the center of each galaxy.  The masked inner region was intended to prevent inaccurate fits resulting from a poorly-fit or variable PSF.  While the default was to allow the sky to vary, our priority was to fix the sky when this failed so as to most closely mimic Blakeslee et al.'s method, and then the last resort was to fix the centers.  For ID 300 (Figure 1), an AGN with a very bright active nucleus (Demarco et al.~\cite{dem05}), none of these fits worked, and so it is left without a one-component fit.   We do not measure the Bumpiness parameter used in Blakeslee et al.~(\cite{bla06}), since visually determining the presence of early-type or late-type features is already an essential part of our method.  Since Blakeslee et al.~found about 80\% of early-type galaxies in their sample to have n $>$ 2.5, we used n $>$ 2.5 plus the presence of visual early-type features to provide a "second opinion" on galaxy type.  A disk galaxy was thus still considered early-type (S0) if the 2-D fit suggested disk domination but the galaxy had n $>$ 2.5 and no spiral structure.  The details of this new method of searching for S0 galaxies are described in the subsection on galaxy classification.

\subsection{Half-Light Radii}
Along with the GALFIT profiles, we also determined half-light radii in the $z_{850}$-band, using Source Extractor results and the IRAF program ELLIPSE.  The Source Extractor FLUX\_ISO parameter was chosen as an estimate of the total flux of the galaxy, since the Source Extractor isophotal area generally corresponded very well with the galaxy's apparent size.  For each galaxy (except for the irregular galaxy 1238b which appears to lack a nucleus), we ran ELLIPSE in IRAF.  The center, ellipticity, and position angle (PA), appropriately rotated to match the different definitions of PA $=$ 0, were provided by Source Extractor.  The semi-major axis was four times the initial half-light radius estimate used in the GALFIT fitting, which was derived from the Source Extractor isophotal area.  This generous initial semi-major axis made sure that the half-light radius would be within the ELLIPSE contours.  To prevent the centers of subsequent ellipses from drifting far from the center of each galaxy, we fixed the ellipse fitting centers (hcenter $=$ yes in the parameter file).  Once we had a full set of concentric ellipses and their semi-major axes, we used INTERPOL in IDL to provide a linear interpolation of the ELLIPSE semi-major axis. The semi-major axis was interpolated at a TFLUX\_E (total flux enclosed by the ellipse) value of one-half of the Source Extractor FLUX\_ISO value.  This interpolated semi-major axis was taken as our half-light radius for the galaxy and used in the final classification scheme to distinguish compact galaxies from other categories.

\subsection{Color Maps}
As in Delgado-Serrano et al.~(\cite{del10}), we created two-band color maps (Zheng et al.~\cite{zhe05}) for each galaxy and its surroundings.  The color maps were created using the $r_{625}$ and $z_{850}$ bands, to provide the best color contrast and most closely mimic the previous studies on field galaxies.  We created two color maps for each galaxy: one that shows the background (unmasked) and one that only shows the parts of any galaxies in the image with flux above the Source Extractor threshold (masked).  To make the maps, the ratio of the $r_{625}$ and $z_{850}$ fluxes was determined pixel-by-pixel from the images.  This ratio image was smoothed with a 2$\times$2 boxcar, and then each pixel of the smoothed ratio image was converted to colors in units of magnitudes.  These color maps were studied to detect the presence of such features as nuclear star-forming regions, supernovae, AGN, normal red nuclei, and dust lanes.  We used color map information in the galaxy classification scheme discussed below.

\subsection{Galaxy classification}

The final galaxy classification scheme is shown in Figure 2, and is primarily based on the decision tree of Delgado-Serrano et al.~(\cite{del10}).  The scheme was modified based on the spatial resolution of our images and the problem, briefly mentioned above, of many galaxies with early-type features having low bulge-to-total (B/T) ratios.  These low B/T ratios may be genuine, or they may be a by-product of overall lower-quality two-component fits.  The galaxies were at slightly higher redshifts and thus of smaller angular size than those of Delgado-Serrano et al.~(\cite{del10}), and were also in more crowded fields.  In Figure 2, we also note the numbers of galaxies excluded from regular E, S0, or Spiral morphology by each criterion.  Seven galaxies were Compact, with half-light radius less than 3 pixels. One galaxy had a high B/T and peculiar or merging features.  Eight galaxies had intermediate B/T and peculiar or merging features.  Sixteen galaxies had low B/T and peculiar or merging features, and six galaxies with low B/T had early-type features with low S\'ersic index.  Seven galaxies had a blue nucleus but no other peculiar features, and two galaxies had a large bulge/disk center separation but no other peculiar features.

Our first modification to the decision tree was to define compact galaxies as having a half-light radius less than 0.15 arcsec or 3 HST/ACS pixels.  This size limit was equivalent to about 1.15 kpc at the cluster distance, or 1.76 times the PSF size.  The change in the physical scale reflects the slightly smaller angular scales at redshift 0.84 vs. redshift 0.6, but is otherwise similar to Delgado-Serrano et al.~(\cite{del10}) at z = 0.6.  The most significant modification was to change the criteria for S0 galaxies to include some disk-dominated objects according to the 2-component fit. The galaxies in question otherwise had all the features of an early-type galaxy: smooth regular structure, a red nucleus, and a S\'ersic index from the one-component fit greater than 2.5. The remaining six disk-dominated galaxies with early-type features that had a one-component S\'ersic index less than 2.5 were then classified as peculiar.  Peculiar features were defined in more or less the same way as in Delgado-Serrano et al.~(\cite{del10}).  These features included a nucleus bluer than the rest of the galaxy (in non-masked color maps), more than 3 pixels of separation between bulge and disk centers, and distorted or irregular visual morphology.  For clarification on the interpretation of color maps in the decision tree, we changed the wording of the ``nucleus redder than outskirts'' criterion to ``nucleus redder (or not bluer) than outskirts.'' This change proved necessary to avoid over-classification of galaxies as Peculiar on the basis of a weak color gradient.  Any of the features discussed above would lead to a galaxy being classified as peculiar.  We also, in a similar manner to widening the criteria for S0 galaxies, widened the criteria for spiral galaxies.  In our system, any galaxy with spiral features, no peculiar features, and a B/T less than 0.8 was classified as a spiral.  Two apparent spiral galaxies shown in Figure 3, ID 328 (left) and ID 306 (right), had a bulge-to-total ratio between 0.5 and 0.8, and one (ID 306) had a one-component S\'ersic index n $>$ 2.5.  Both still had a red nucleus, non-distorted spiral features, and a center separation of less than 3 pixels, and so we believe that classifying such objects as Peculiar would not be warranted.  ID 328 could also alternatively be classified as S0, but since its small disk shows hints of structure, it is classified as a spiral by J. N.  The frequency of such debatable classifications will be discussed later in this section. 

\begin{figure}
\centering
\includegraphics[width=8.5cm]{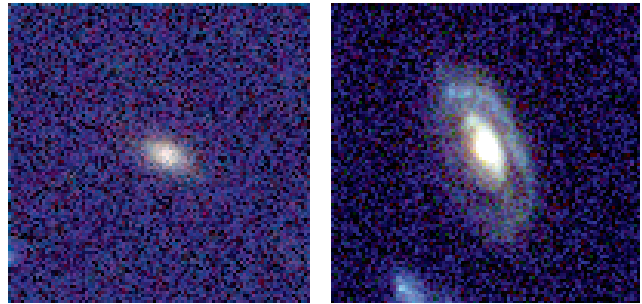}
\caption{5 arcsec by 5 arcsec color images of ID 328 (left) and ID 306 (right), two galaxies of spiral appearance whose two-component fits assigned more than 50\% of light to the bulge.  Galaxy 306 also had a one-component profile fit with a S\'ersic index above 2.5, but its clear arm structure makes it better classified as a spiral than an S0.  For this reason we allow apparently bulge-rich galaxies with regular arm structure and red nuclei to still be considered spiral galaxies.  Background levels and intensity scale were adjusted to show the structure of these galaxies' disks.}
\label{FigBulgeSp}
\end{figure}

\begin{figure}
\centering
\includegraphics[width=8.5cm]{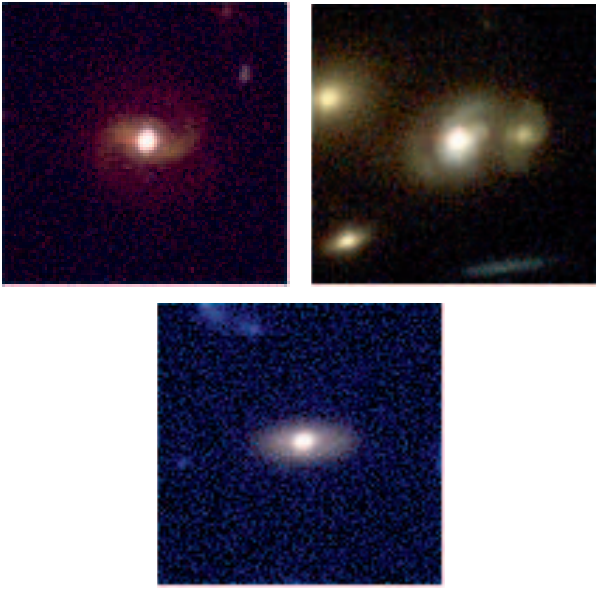}
\caption{5 arcsec by 5 arcsec color images of three galaxies assigned to our Peculiar, Merging, or Compact categories that do not easily fall into the traditional sub-categories of this group.  Upper left: ID 300, an AGN excluded from the regular galaxy category due to a blue nucleus (see text).  Upper right: ID 557, another AGN (Demarco et al.~\cite{dem05}) excluded due to a blue nucleus, although it may also be merging with its neighbor to the right.  Bottom: ID 258, an S0/Sa-like galaxy considered ``compact'' in our analysis due to its high-surface-brightness nucleus.  All images have bright nuclei, and so background levels and intensity scales were adjusted to show both the bright core and the much fainter surrounding features.}
\label{FigPecOther}
\end{figure}

\begin{figure}
\centering
\includegraphics[width=8.5cm]{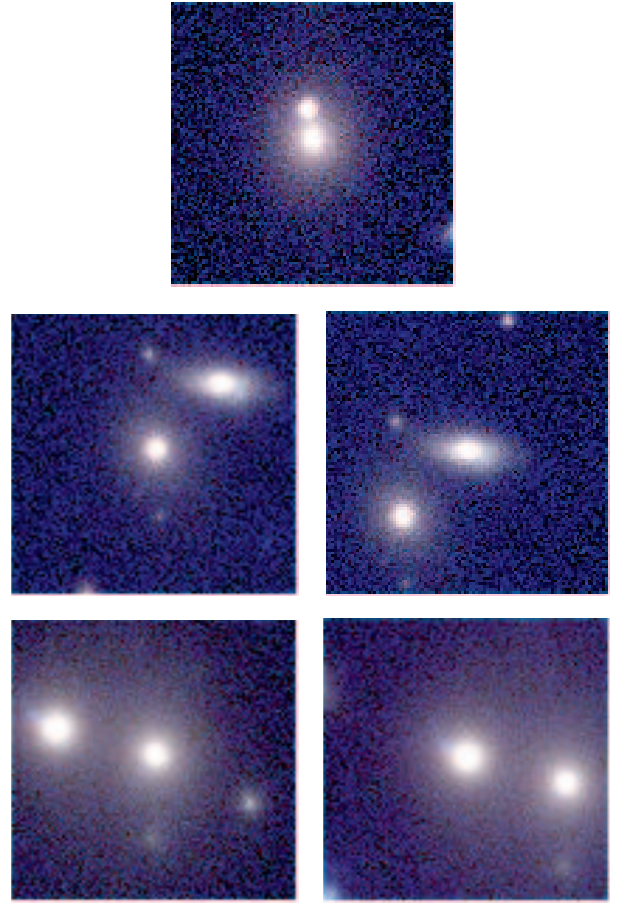}
\caption{Examples illustrating our criteria for assigning or refusing to assign a close visual pair of galaxies as a merger.  The top image is ID 387, assigned as a merger because its secondary core is too close to the primary core to have been registered as a separate galaxy, and appears to lack its own undistorted halo.  The center images are IDs 131 (left) and 1501 (right), two closely associated but otherwise distinct-looking early type galaxies.  The bottom images show IDs 1466 (left) and 1467 (right), also with well-separated cores and mostly independent envelopes.  The blue streak emerging from the core of ID 1467 may be a jet or lensed background galaxy.  All images are 5 arcsec by 5 arcsec.}
\label{MergNonmerg}
\end{figure}

We did not further classify peculiar or merging galaxies in the schematic shown in Figure 2, since we do not concern ourselves with these distinctions in our analysis of morphological types.  Most peculiar, compact, or merging galaxies fit the Pec/T (tadpole), Pec/M (possible merger remnant), Pec/Irr (late type/Magellanic irregulars), M (active merger), and Compact (C) categories from Delgado-Serrano et al.~(\cite{del10}).  However, a few galaxies such as those in Figure 4 did not easily fit into the above categories.  Seven galaxies, including the two AGN shown in the top of Figure 4, were considered peculiar purely on the basis of a blue nucleus.  Exclusion from regular morphology on the basis of a blue nucleus is a rather conservative criterion to use at intermediate-to-high redshift, in which disk galaxies are likely to be considerably less evolved than in the local universe.  Despite this fact, we retain this criterion in order to be consistent with the methodology of Delgado-Serrano et al.~(\cite{del10}).  Only 18\% of galaxies in the peculiar sample, including these two AGN, met only this criterion of peculiarity.  We also did not distinguish in our decision tree between visual ``E features'' and ``S0 features,'' as such a distinction can be very difficult and is often made heavily on the basis of inclination.  We therefore insisted on a B/T $>$ 0.8 in combination with early-type features to classify a galaxy as Elliptical.  Any other regular early-type galaxy, with a B/T above 0.5 and/or a fitted 1-D S\'ersic index above 2.5, was thus classified as S0.  If the galaxy visually appeared to be elliptical but did not have a B/T greater than 0.8, we assumed it to be a face-on S0 in this classification system.

The fitting and final classifications of galaxies were performed by J. N.  Two other team members (H. F. and R. D.) also performed classifications to test the uncertainty. The R. D. classifications were performed for the full sample, in which there were eight galaxies (6.5\%) for which assignment to the E, S0, Sp, or Pec/C/M categories differed from that of J. N.  In the H. F. classifications, in which 50 galaxies were classified, disagreement with J. N. was found for about eight of the 50 galaxies (16\%).  Averaging these figures gives about 11\% uncertainty in the morphological classifications.

Ultimately, we decided that the Pec/M or Merger morphology would not be assigned to close pairs with two separate galaxy IDs, or which appeared to have separate or mostly separate undistorted envelopes.  Assigning such close but distinct pairs as mergers contributed to the relatively high disagreement rate between the J. N. and H. F. classifications.  ID 387 (top of Figure 5), the core galaxy of the southern sub-cluster (Demarco et al.~\cite{dem05}), was counted as a merger since its second core lies very close to the main galaxy and does not have a separate ID.  Galaxies 131 and 1501 (middle of Figure 5), on the other hand, were considered two separate regular early-types by J. N. despite being visually closely associated.  They have two separate galaxy IDs and mostly distinct haloes.  By the same criteria, the two central galaxies of the northern sub-cluster (Demarco et al.~\cite{dem05}), 1466 and 1467 (bottom of Figure 5), were also counted as two separate early-types rather than a merger.  The blue streak in 1467 is very small, narrow, and close to the nucleus, and so is thought to probably be either a jet or a lensed background galaxy rather than a cannibalized satellite.  While the latter two pairs of galaxies may merge by z = 0, at z = 0.84 they appear to be tidally undistorted and mostly separate entities.

\section{Results}

\begin{figure*}
\centering
\includegraphics[width=18cm]{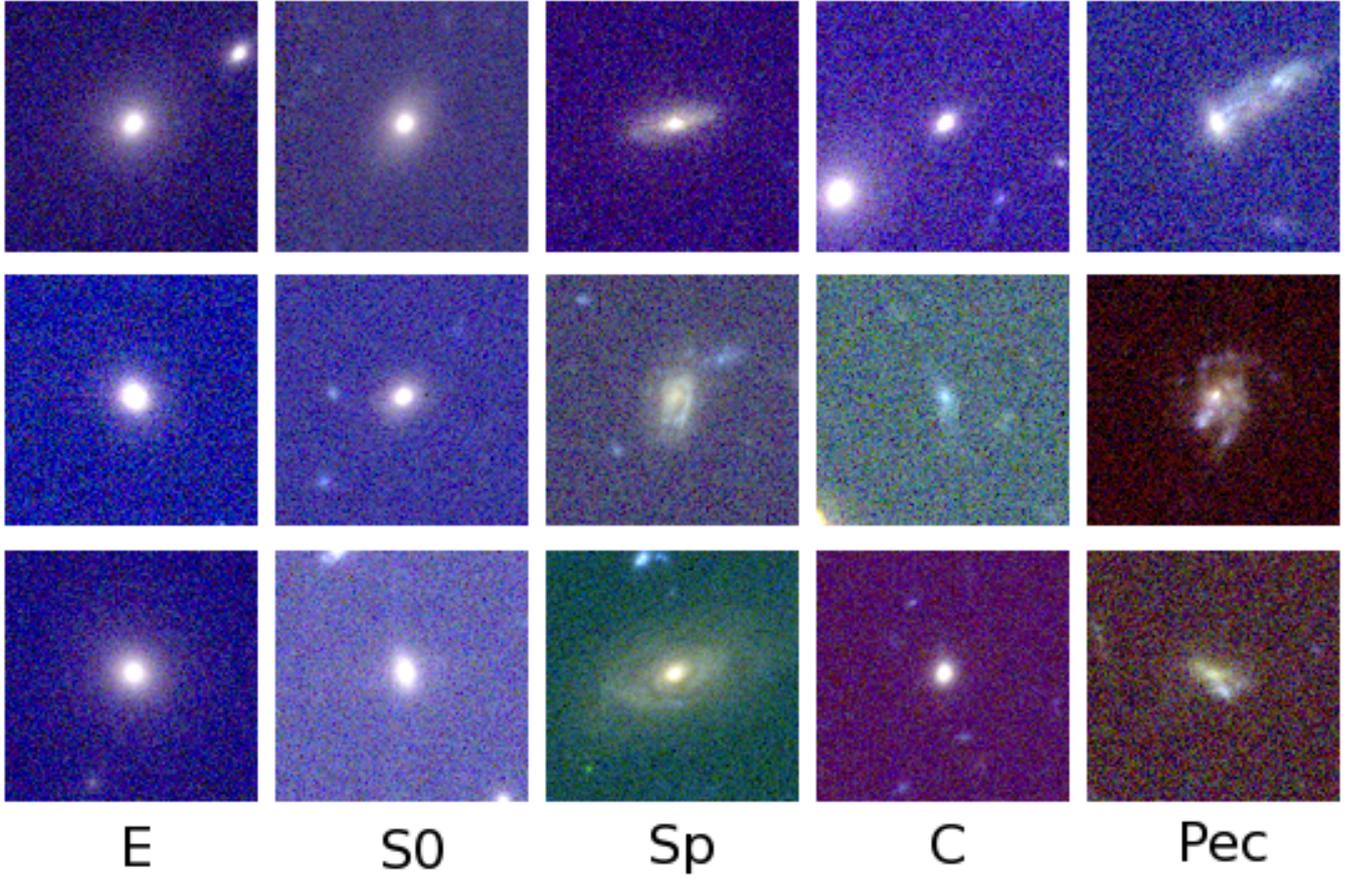}
\caption{Examples of each type of galaxy.  The far left column shows elliptical galaxies 85 (top), 344 (middle), and 468 (bottom).  The second column to the left shows S0 galaxies 332 (top), 81 (middle), and 248 (bottom).  The central column shows spiral galaxies 18 (top), 47 (middle), and 204 (bottom).  The second column to the right shows compact galaxies 129 (top), 145 (middle), and 511 (bottom).  The far right column shows non-compact peculiar galaxies 377 (top, Tadpole features), 522 (middle, Irregular features), and 1112 (bottom, Merger features).  All images are 5 arcsec by 5 arcsec.}
\label{FigExemp}
\end{figure*}

\begin{table*}
\caption{\label{t2}Galaxy Type Fractions by Morphology}
\centering
\begin{tabular}{lccccc}
\hline\hline
Sample & Frac. E & Frac. S0 & Frac. Sp & Frac. Pec\&M & Frac. C\\ 
\hline      
Full sample & 0.11$\pm$0.03 & 0.36$\pm$0.05 & 0.15$\pm$0.03 & 0.32$\pm$0.05 & 0.06$\pm$0.02\\
LDMD & 0.05$\pm$0.03 & 0.23$\pm$0.06 & 0.19$\pm$0.06 & 0.48$\pm$0.09 & 0.05$\pm$0.03\\
IDMD & 0.16$\pm$0.06 & 0.45$\pm$0.11 & 0.16$\pm$0.06 & 0.13$\pm$0.06 & {\bf{0.10$\pm$0.05}}\\
HDMD & 0.21$\pm$0.09 & 0.58$\pm$0.16 & 0.00$\pm$0.00 & 0.21$\pm$0.09 & 0.00$\pm$0.00\\
High mass & 0.21$\pm$0.07 & 0.49$\pm$0.12 & 0.12$\pm$0.05 & 0.16$\pm$0.06 & 0.02$\pm$0.02\\
Low mass & 0.03$\pm$0.02 & 0.28$\pm$0.07 & 0.15$\pm$0.05 & 0.44$\pm$0.09 & 0.10$\pm$0.04\\
No mass estimate & 0.08$\pm$0.08 & 0.25$\pm$0.14 & 0.25$\pm$0.14 & 0.42$\pm$0.19 & 0.00$\pm$0.00\\
High mass, LDMD & 0.13$\pm$0.09 & {\bf{0.34}}$\pm$0.15 & 0.20$\pm$0.12 & 0.33$\pm$0.15 & 0.00$\pm$0.00\\
Low mass, H/IDMD & 0.04$\pm$0.04 & 0.44$\pm$0.13 & 0.12$\pm$0.07 & 0.28$\pm$0.11 & 0.12$\pm$0.07\\
\hline                  
\end{tabular}

\end{table*}

Table 2 shows the type fractions of each morphological category of galaxies (E, S0, Spiral, Peculiar/Merger, and Compact) for various samples within the cluster J0152.7, starting with the full sample of 124 galaxies.  For consistency with Delgado-Serrano et al.~(\cite{del10}), we count the Compact galaxies among the Peculiar objects in our later analysis of type fractions.  Some Compact galaxies, however, do not appear particularly Peculiar in morphology: see, for example, ID 258 in Figure 4, which appears S0 or Sa with a compact bulge.  Figure 6 shows three exemplars each of Elliptical, S0, Spiral, Compact, and non-compact Peculiar galaxies.  The high dark-matter density (HDMD), intermediate dark-matter density (IDMD), and low dark-matter density (LDMD) subsamples represent high, intermediate, and low density regions, defined as in Demarco et al.~(\cite{dem10}).  These density region categories are based on the Jee et al.~(\cite{jee05}) weak-lensing maps.  These and other subsamples will be discussed in the appropriate subsections.  In the full sample, we classified a total of 59 galaxies as regular early types, filling all our necessary criteria for E or S0 galaxies.  These constitute 47\% of the total cluster sample, about 2.8 times the Delgado-Serrano et al.~(\cite{del10}) early type fraction for the field.  This early-type fraction is much lower than in the mass-selected samples of Holden et al.~(\cite{hol07}) and van der Wel et al.~(\cite{van07}).  It is, however, consistent with other z $\sim$ 1 cluster studies such as Mei et al.~(\cite{mei12}) at z = 1.3 and van Dokkum et al.~(\cite{vando00}) at z = 0.83.  Both find early-type fractions under 50\% in clusters.  These latter two studies, however, used one- and two-band imaging, respectively, in blue rest-frame colors, and so may have classification errors resulting from insufficient color information.  Of our regular early-type galaxies, 45 are classified as S0, and 14 are classified as Elliptical.  Eighteen galaxies (15\% of the total sample) are classified as spirals.  Of the remaining galaxies, seven are Compact and 40 are non-compact Peculiars and mergers, making 47 peculiar, compact, and merging galaxies altogether, or 38\% of the total sample.  Both our results and the studies with higher early-type fractions are consistent with Jaff\'e et al.~(\cite{jaf11}), who find an excess of kinematically disturbed but morphologically regular galaxies in cluster infall regions at z $\sim$ 1.  In the full sample of spectroscopic cluster members, this means that many galaxies have yet to evolve into morphologically regular early types with normal color gradients.  However, the typical evolutionary state of the galaxies is highly dependent on location and stellar mass, as we will discuss in the subsections below.

\subsection{Morphology vs. environment within the cluster and comparison to the field}

\begin{figure}
\centering
\includegraphics[width=8.5cm]{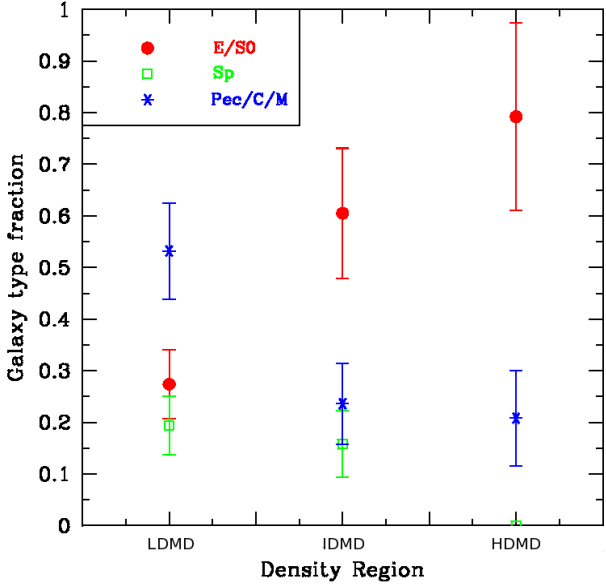}
\caption{Galaxy type fraction vs.~environmental density region (HDMD= high dark matter density, IDMD = intermediate dark matter density, LDMD = low dark matter density) for J0152.7 cluster galaxies.   Error bars in galaxy type fractions are from Poisson counting statistics.  There is a strong increase in regular Elliptical and S0 galaxies and decrease in Peculiar, Compact, and Merging galaxies in the intermediate-density and high-density regions compared to the cluster outskirts.  This pattern suggests that large-scale morphological transformation may occur before reaching the cluster core.}
\label{FracvsEnv}
\end{figure}

In Demarco et al.~(\cite{dem10}), the spectra of various galaxies in J0152.7 were stacked and compared according to regions of high, intermediate, and low dark-matter density as determined by the weak lensing map of Jee et al.~(\cite{jee05}).   This allows one to understand the correlation between the properties of galaxies and environmental density.  The latter can affect the frequency, severity, and type of various galaxy-transforming processes such as mergers, harassment, and ram pressure.  We expect these processes not just to affect the star formation history and current star formation rates as studied in Demarco et al.~(\cite{dem10}), but also galaxy morphology.  By comparing the morphologies of galaxies in J0152.7 based on the local dark-matter density, we may be able to determine the stage of cluster infall at which galaxies experience the bulk of their transformation.   This, in turn, may help determine which processes play the greatest roles in the transformation from late-type, peculiar, compact, or merging systems to regular early-types.  Demarco et al.~(\cite{dem10}) created their density region map by translating the $\kappa$ ($\Sigma/\Sigma_{c}$) values of Jee et al.~(\cite{jee05}) into dark-matter densities.  The high-density regions were defined as more than 20 $\Sigma_{DM}$, intermediate density regions as 5 to 20 $\Sigma_{DM}$, and low-density regions as $<$ 5 $\Sigma_{DM}$.  In these definitions, $\Sigma_{DM}$ = 0.0057 × $\Sigma_{c}$, and $\Sigma_{c}$ is the critical density of 3650 M$_{\odot}$ pc$^{-2}$ from Blakeslee et al.~(\cite{bla06}).  Figure 7 shows the distribution of early (E/S0), spiral, and peculiar galaxy type fractions as a function of density region.

Half of the 124 spectroscopically confirmed galaxies in the ACS imaging lie in the lowest-density outskirts of the cluster (LDMD in Table 2 and Figure 7), outside the merging cluster cores or infalling rich groups.  Among these galaxies, the fraction of regular E/S0 morphologies is 27\%, or 17 out of 62 galaxies, about 1.5 times higher than the early type fraction found for the field at similar redshifts in Delgado-Serrano et al.~(\cite{del10}).  This is consistent with the findings of Balogh et al.~(\cite{bal11}), in which galaxies in groups at z $\sim$ 0.9 were redder and earlier-type than more isolated field galaxies at the same redshift. Our spiral fraction in the low-density regions is about 19\%, more similar to the Delgado-Serrano et al.~(\cite{del10}) field early-type fraction (17\%) than to the field spiral fraction (31\%).  This result suggests that the cluster outskirts might already be biasing the formation of peculiar galaxies toward becoming early types rather than spirals.  The peculiar, compact, and merger fraction is 53\%, about the same as the Delgado-Serrano et al.~(\cite{del10}) intermediate-redshift field.

Most of the difference between our cluster outskirts and Delgado-Serrano et al.~(\cite{del10}) is made up by S0s (See Table 2 for the S0 vs.~E fractions), for which we allow a ``second opinion'' based on the one-component S\'ersic index.  This difference may reflect our different criteria for the S0 vs.~spiral distinction.  If we exclude the four ``second opinion'' S0s and classify them as spirals, we would reduce the early-type fraction to 21\% and increase the spiral fraction to 26\%, only a small change similar to the magnitude of the uncertainties.  However, in order to meet our S0 criteria, the disk-dominated galaxies had to lack spiral arms and have a high S\'ersic index in a single-component fit.  Galaxies meeting these criteria are probably not best classified as pristine spirals.  In a visual-only classification, these galaxies would likely be classified as elliptical or S0.   

The intermediate-density regions (IDMD in Table 2 and Figure 7) house 38 of the confirmed cluster members. The IDMD includes three types of environments.  One such environment is the fringe around the merging cluster cores identified in Demarco et al.~(\cite{dem05}).  Another is a pair of rich infall groups to the south and east of the two main merging sub-clusters (Demarco et al.~\cite{dem05}, \cite{dem10}).  The third component of the IDMD consists of a few isolated dense spots in the dark matter maps. The IDMD galaxies have over twice the early-type fraction of the low-density regions, at 61\% (23 out of 38 objects).   The spiral fraction in the intermediate density regions is only slightly lower than in the outskirts, at 16\% (6 out of 38 objects).  The peculiar fraction is hardly more than half that found in the outskirts, at 23\%  (9 out of 38 galaxies).  This suggests that early-type galaxies, in the studied cluster, can either form from spiral galaxies or directly from peculiar galaxies.  And perhaps the spiral fraction in the outskirts is diminished not because spirals are evolving into early types, but because more peculiar galaxies than in the field became early types instead of spirals.

\begin{figure}
\centering
\includegraphics[width=8.5cm]{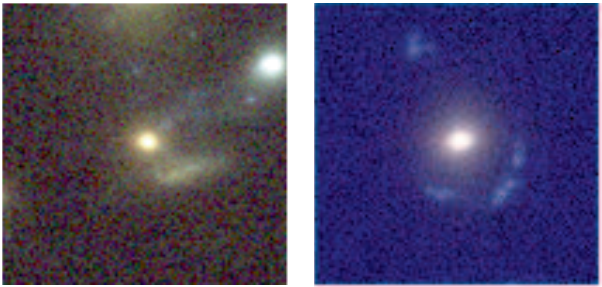}
\caption{Left image: Galaxy 1454, one of the ``peculiar/merging'' galaxies identified in the high-density regions, classified as Pec/M for its apparent interaction with a disk-like system less than 1 arcsec from its core and faint blue haze which may be tidal debris from an interaction.  Right image: Galaxy 1278, a regular early type surrounded by lensed images of a background galaxy.  Note the difference in color and morphology between the very blue, curved lensed images surrounding Galaxy 1278, vs. the yellow, only slightly distorted disk companion of Galaxy 1454.  These differences suggest that the latter is more likely an interaction than a case of lensing or chance superposition.  Both images are 5 arcsec by 5 arcsec.}
\label{MergervsLens}
\end{figure}

The high-density regions (HDMD in Table 2 and Figure 7), consisting of the two merging cluster cores themselves, have a morphological profile typical of a mature galaxy cluster.  The E/S0 fraction in the HDMD is 79\%  (19 out of 24 galaxies).  No spirals were found in the high-density regions, and five galaxies in these inner regions were classified as peculiar, giving a peculiar fraction of 21\%. ID 557 (Figure 4, upper right) is an AGN that may be interacting with a neighbor.  ID 387 (Figure 5, top), the central galaxy of the southern cluster, has a second high-surface-brightness core about 0.6 arcsec from the main core and so is classified as a merger.  ID 1454 (Figure 8, left) has a high-surface-brightness core, an faint blue trail extending northwest, and a small disk or peculiar galaxy less than 1 arcsec from its core with which it may be merging.  The companion of ID 1454 could be a gravitational arc, but it is not strongly curved, and its color contrast with the main galaxy is less than in more obvious cases of arcs (e.g., ID 1278, right side of Figure 8).  The brightest regions of the companion of ID 1454, in fact, seem to have a color only slightly bluer than the center of ID 1454 itself.  The remaining two ``peculiar'' galaxies in the high-density regions are S0-like in appearance.  They have no signs of spiral arms, but have very low bulge-to-total ratios and S\'ersic indices and thus cannot be officially classified as either S0s or spirals.

As Figure 7 shows, with increasing local dark-matter density and proximity to a cluster core, the early type fraction increases.  The early type fraction grows dramatically between the LDMD and IDMD regions.  The peculiar fraction is more rapidly diminished than the spiral fraction between these regions, while the spirals primarily disappear going from the IDMD to HDMD regions.  Thus, the intermediate-density regions --- massive rich infall groups and the immediate outskirts of the cluster core --- seem to be an especially rich and active site of morphological evolution in the cluster environment.  However, the Poisson counting errors lead to rather large uncertainties in the early-type fractions of the high and intermediate density regions.  The difference in early-type fraction between the intermediate and high density regions may thus be relatively modest.  Also, depending on the orbit of the galaxy, the region in which the bulk of the transformation occurred is not necessarily that in which the galaxy is currently found.  Galaxies in the intermediate-density regions on the fringes of the main sub-clusters (as opposed to in the eastern or southern infall groups) may have already crossed the cluster cores.

In the cluster environment, there appear to be two possible evolutionary tracks that late-type galaxies can take.  One path is evolution from peculiar galaxies into spirals at higher redshift and then from spirals into early types, as appears to be the case going from intermediate to high density.  The other is evolution from a peculiar, compact, or merging system directly into an early type without ever having become a regular spiral beforehand.   Our results are consistent with those of Prieto et al.~(\cite{pri13}) in which early-type galaxies appear to form from the evolution of irregular disks into {\it{irregular}} spheroids, and then from irregular spheroids into regular spheroids.  The studies of Delgado-Serrano et al.~(\cite{del10}) and Flores et al.~(\cite{flo06}) suggest that many field disk galaxies at z $\sim$ 0.8 have yet to evolve into regular spirals.  Falling into a cluster environment may change the fate of these systems, such that they evolve into regular elliptical or S0 galaxies instead of regular spirals.  It is possible that stellar mass may influence these evolutionary tracks.  For instance, low stellar mass objects may be more likely to directly evolve from peculiar into elliptical or lenticular, while higher stellar mass objects may first pass through a regular spiral phase.  It is also possible that low-mass peculiar galaxies are destroyed by passing through the cluster core.  In the next subsection, we will further discuss morphology as related to stellar mass in the cluster.

\subsection{Stellar mass}

\begin{figure*}
\centering
\includegraphics[width=12cm]{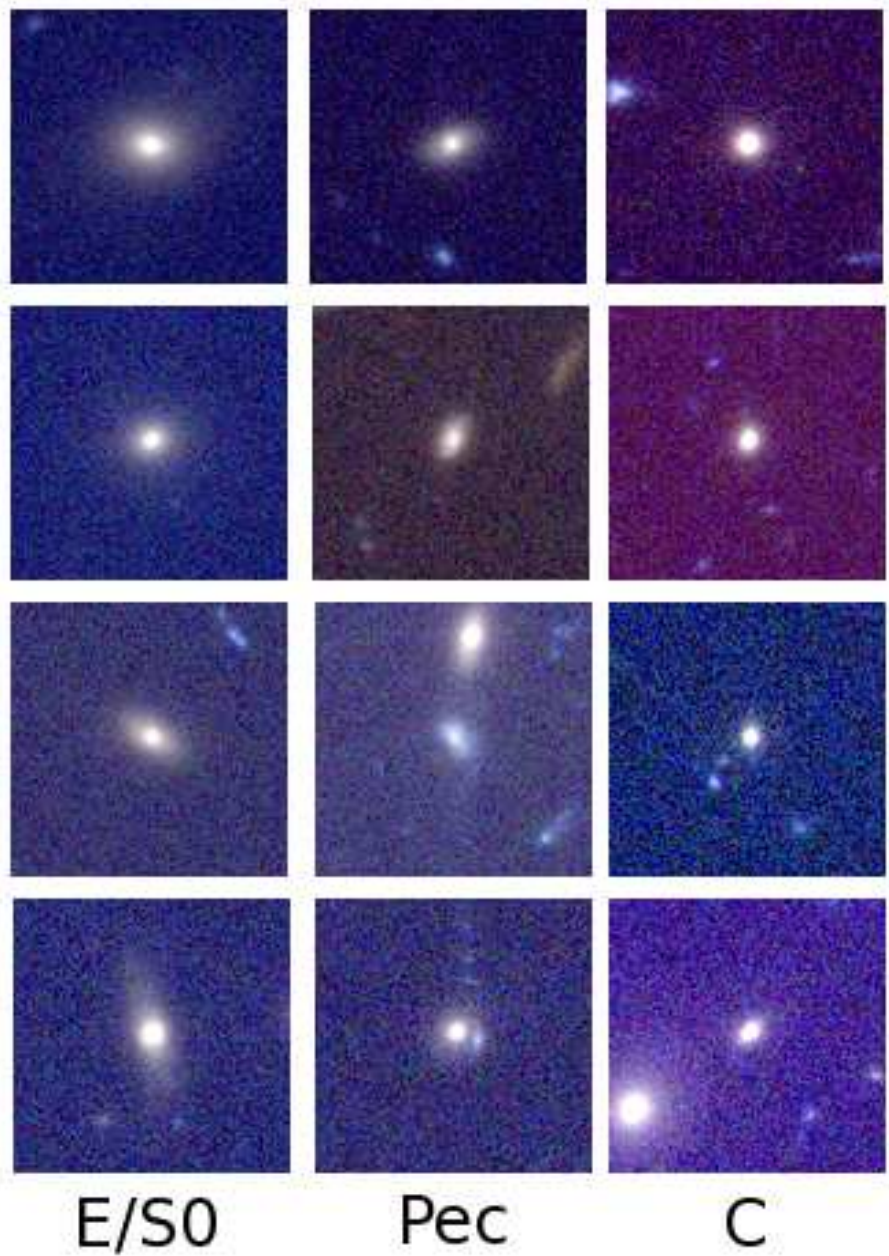}
\caption{Examples of galaxies identified as early-types (E/S0, T-types -5 to -1) by Postman et al.~(2005) and how we classified them.  The left column shows examples of galaxies that are also identified as early types in our scheme.  The top two galaxies of the left column (1172 and 1184) are elliptical and the bottom two (397 and 547) are S0.  The middle column shows galaxies that we classify as Peculiar: 626 (too disky and too low S\'ersic index to be S0), 1146 (possible tidal distortions), 898 (irregular), and 1132 (possible merger).  The right column shows galaxies we classify as Compact: 347, 511, 1067, and 129.  (All examples shown resemble early types but lack a disk or halo.)  All images are 5 arcsec by 5 arcsec.}
\label{FigPostEarly}
\end{figure*}

Our 124 cluster galaxies all have morphological T-types, mostly from Postman et al.~(\cite{pos05}).  The Postman et al.~definition of early type galaxies is T $\leq$ -1, including E, S0, and S0/a galaxies.  This system ignores most of what we label as Peculiar features in such galaxies, such as tidal distortions, nuclei bluer than the outskirts, compactness, or potential merger activity.  (Only Magellanic-type irregular features are identified in the T-types.)  Examples of early-T-type galaxies from Postman et al.~(\cite{pos05}) and our classifications of them are shown in Figure 9.  The left column shows cases where we agreed with Postman et al. in assigning these galaxies to E or S0 types, and the other columns show cases of disagreement.  Among our 124 galaxies, about 58\% of the sample (72 galaxies) are early types according to the Postman et al.~(\cite{pos05}) T-type classification scheme.  Of these 72 early-T-type galaxies, 54 are also early-type according to our classification scheme, like the objects in the left column of Figure 9.  One object is classified as a spiral.  Twelve objects are classified as peculiar or merging including heavily disk-dominated S0-like galaxies (middle column in Figure 9), and five are classified as Compact (right column in Figure 9).  Most of the galaxies that do not currently meet our regular early-type criteria, especially disk-dominated S0-like objects and early-type mergers, may be in the process of evolving into regular early types.

To explore the effects of stellar mass, we broke down our sample into high-mass galaxies (M $>$ 3.98 $\times$ 10$^{10}$ M$_{\sun}$; Holden et al.~\cite{hol07}), low-mass galaxies (M $<$ 3.98 $\times$ 10$^{10}$ M$_{\sun}$), and galaxies with no mass estimates due to incomplete photometry.  Holden et al.~(\cite{hol07}) chose their lower mass limit of 3.98 $\times$ 10$^{10}$ M$_{\sun}$ (10$^{10.6}$ M$_{\sun}$) because they found their redshift and/or morphological samples incomplete below this limit.  The fractions of each galaxy type for each of the above three subsamples are shown in the fifth, sixth, and seventh rows of Table 2.  The galaxy type fractions for the high-mass and low-mass samples are plotted in Figure 10.  Above the Holden et al.~(\cite{hol07}) limit, 36 of the 51 massive galaxies (70\% of the total massive sample) are early-type.  This is still lower than the typical early type fraction of 80\% that Holden et al.~(\cite{hol07}) found among massive cluster galaxies using the Postman et al.~(\cite{pos05}) T-types, but higher than our total 47\% early-type fraction for the entire cluster sample.  The spiral and peculiar/compact/merger fractions for our ``massive'' subsample according to our scheme are 12\% and 18\% respectively.  Massive galaxies are thus more likely to be regular early types, and less likely to be peculiar, compact, or merging, than galaxies in the cluster as a whole.  The low-mass galaxies --- those with an estimated stellar mass below 3.98 $\times$ 10$^{10}$ M$_{\sun}$ --- have a much lower early-type fraction and higher peculiar fraction.  The early type fraction for low-mass galaxies is only 31\%, the spiral fraction is 15\%, and the peculiar, compact, and merging fraction is 54\%, similar to the cluster outskirts.

\begin{figure}
\centering
\includegraphics[width=8.5cm]{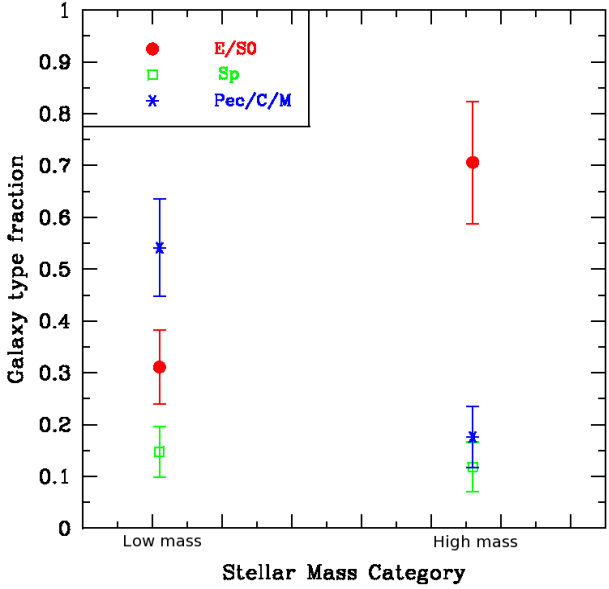}
\caption{Galaxy type fraction vs.~logarithm of mean sample mass for two stellar-mass-selected samples of J0152.7 cluster galaxies: those above and below 3.98 $\times$ 10$^{10}$ solar masses (Holden et al.~\cite{hol07}).   Error bars in galaxy type fractions are from Poisson counting statistics.  More massive galaxies are considerably more likely to be E or S0 and less likely to be peculiar, compact, or merging than are low-mass galaxies.}
\label{FracvsMag}
\end{figure}

\begin{figure}
\centering
\includegraphics[width=8.5cm]{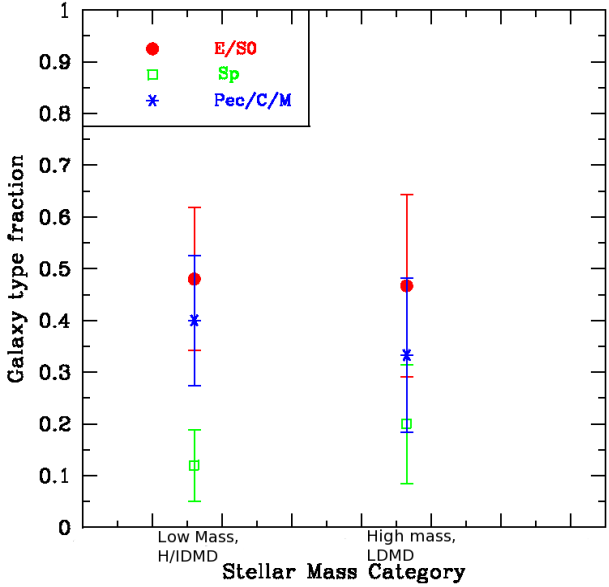}
\caption{Galaxy type fraction vs.~logarithm of mean sample mass for two samples of J0152.7 cluster galaxies that are ``atypical'' in the light of the correlation of galaxy mass with environment. Low-mass galaxies in regions of high to intermediate density are shown on the left, and high-mass galaxies in low-density outskirts are shown on the right.   Error bars in galaxy type fractions are from Poisson counting statistics.  Galaxy type fractions are similar among massive galaxies in low-density regions and low-mass galaxies in high to intermediate density regions.  This suggests that mass and environment both affect galaxy morphology.}
\label{FracAtyp}
\end{figure}

The statistics for high-mass vs. low-mass galaxies are similar to those in the high/intermediate and low dark matter density regions, respectively.  This finding suggests a possible degeneracy or correlation between a galaxy's stellar mass and its environment, which we do find in our sample.  About 75\% of the galaxies in the high-density regions are more massive than 3.98 $\times$ 10$^{10}$ M$_{\sun}$.  Only 47\% of those in the intermediate-density regions and 24\% of those in low-density regions are above this mass limit.  This may be attributed to mass segregation within the cluster, in which the more massive galaxies migrate toward the cluster center (e.g., Pracy et al.~\cite{pra05}; Zandivarez \& Mart\'inez \cite{zan11}).

We can partially break the degeneracy caused by the stellar mass-environment correlation by looking specifically at high-mass galaxies in the low-density regions, and low-mass galaxies in high to intermediate density regions.  Figure 11 and the last two rows of Table 2 show the galaxy type fractions for these ``atypical'' subsamples.  The 15 massive galaxies in low-density regions are more regular and early-type than the average for low-density regions.  Of these massive LDMD galaxies, 47\% are regular early types, 20\% are regular spirals, and 33\% are peculiar, compact, or merging.  These massive galaxies have a lower early-type fraction than the populations of intermediate and high density regions, suggesting that environment may play a role even in the evolution of massive galaxies.  The 25 low-mass galaxies in the high and intermediate density regions also have intermediate morphology statistics. Among low-mass, high- to intermediate-density galaxies, 48\% are regular early types (mostly S0), 12\% are regular spirals, and 40\% are peculiar, compact, or merging.  Within the Poisson counting uncertainties, there is not even a marginally significant ($> 1 \sigma$) difference in the early-type, spiral, and peculiar fractions between either of these ``atypical'' samples.  Thus, it appears that both environmental density and galaxy stellar mass may affect galaxy evolution.  

Several studies (e.g. Gobat et al.~\cite{gob08}; Rettura et al.~\cite{ret10}, \cite{ret11}; Muzzin et al.~\cite{muz12}) suggest that a rich environment could stop star formation prematurely, particularly in intermediate to low-mass galaxies.  A premature end to star formation in galaxies falling into the cluster may convert irregular or spiral galaxies into early types as star formation in arms and clumps ceases.  Internal dynamics of the galaxy may also contribute to transformation, as will be discussed in the paragraph below.  To be counted as S0s or Ellipticals in our classification, these former spiral and irregular galaxies would need to have a high S\'ersic index or bulge domination by the time their visible structure faded.  

We found a total of six galaxies in the cluster, officially classified as Peculiar, that were disk-dominated with a low S\'ersic index but an S0-like appearance (e.g., object 626 in Figure 9).  These galaxies could represent transitional objects, which would have been spirals or Magellanic irregulars were it not for a premature end to their star formation.  Three of them (474, 682, and 5034) are low-mass galaxies in the high or intermediate density regions, with typical passive spectroscopic features.  The other three (626, 928, and 1483) are located in the outskirts and, unlike their counterparts in the denser regions, have E+A features (notable H$\delta$ and H6 lines), suggesting recent quenching.

To gain some insight on what might transform the morphology of our ``near-S0'' transitional galaxies from an initial spiral or Magellanic irregular appearance, we can compare the dynamical friction timescale with the cluster crossing time.  Dynamical friction can convert star-forming clumps in a galactic disk into a bulge, while a cluster crossing would expose the galaxy to ram pressure and/or harassment.  Our test object for this comparison is Galaxy 626, a low-mass (2.1 $\times$ 10$^{10}$ M$_{\sun}$) transitional galaxy in the cluster outskirts.  We use the dynamical friction timescale equation from Noguchi (\cite{nog00}, \cite{nog01}).  We assume a gas mass of 4 $\times$ 10$^{9}$ M$_{\sun}$, a total galaxy mass (including dark matter) of 5 $\times$ 10$^{10}$ M$_{\sun}$, a typical clump mass of 2.1 $\times$ 10$^{8}$ M$_{\sun}$, and a total radius of about 5.34 Kpc.  (Our total radius estimate is nearly three half-light radii of Galaxy 626).  With these numbers, we find a dynamical friction timescale of about 0.34 Gyr.  To estimate the crossing time of Galaxy 626, we use the equation of Sarazin (\cite{sar88}).  Galaxy 626 lies at 705 kpc from the cluster center, and is assumed to travel at $\sigma_v$ = 1300 km s$^{-1}$ (Girardi et al.~\cite{gir05}; Demarco et al.~\cite{dem10}).  This gives a crossing time of about 0.54 Gyr.  The estimated dynamical friction timescale in Galaxy 626 is shorter than its estimated crossing time.  Thus, Galaxy 626 and other transition galaxies may be self-transforming even before they spend large amounts of time in the denser regions of the cluster.

\subsection{Comparison to MS 1054}

Of the morphological studies done on galaxy clusters at similar redshift to J0152.7, that of van Dokkum et al.~(\cite{vando00}) for the dynamically young cluster MS 1054-03, at z = 0.83, is perhaps the study most similar to ours.  It has a luminosity-selected galaxy sample in a dynamically young cluster, and a refined visual galaxy classification system that accounts for mergers and merger remnants.  This study provides the most similar early-type fraction to ours among cluster studies at z $\sim$ 0.8.  They found only 44\% of the galaxies in MS 1054-03 to be undisturbed early types, very similar to our 47\% of J0152.7 members.  Both of these early type fractions are also notably lower than in the Postman et al.~(\cite{pos05}) and Holden et al.~(\cite{hol07}) estimates.  The difference from the former study reflects the distinction between undisturbed and merging galaxies with otherwise early-type morphology.  The difference from the latter study indicates both the undisturbed vs.~merging distinction and luminosity sampling vs.~mass sampling.  The non-early-type categories of van Dokkum et al.~(\cite{vando00}) are spirals (39\% of MS 1054-03 members) and mergers-in-progress (17\% of MS 1054-03 members).  Their spiral category is probably more inclusive than our undisturbed spiral category (15\% of J0152.7 members), and their merging category is more exclusive than our peculiar, compact, and merging category (38\% of J0152.7 members).  Still, the overall fraction of galaxies that do not appear to be regular early types is similar between the two classification schemes.  This suggests similar states of galaxy evolution in these two z $\sim$ 0.8, dynamically young clusters.  However, it should be noted that the van Dokkum et al.~(\cite{vando00}) study, while distinguishing mergers from regular galaxies like our study, did not involve profile fitting or the use of colors redder than rest-frame $B$.

\section{Conclusions}
We performed a new morphological study of 124 spectroscopically confirmed members of the galaxy cluster RX J0152.7-1357.  We used a modified version of a classification scheme developed for field galaxies by Neichel et al.~(\cite{nei08}) and Delgado-Serrano et al.~(\cite{del10}), which correlated details of profiles, colors, and shapes with the kinematics of galaxies.  From our analysis of 124 spectroscopically confirmed members of J0152.7, we found 47\% of the galaxies to be regular, passive early-types.  This early-type fraction is about 2.8 times the overall early-type fraction found in the field at intermediate redshift in Delgado-Serrano et al.~(\cite{del10}). We broke down the morphology of the galaxies by local dark-matter density as determined by the Jee et al.~(\cite{jee05}) weak-lensing maps to explore the cluster's internal morphology-density relation.  We found an especially drastic drop in the fraction of peculiar, compact, and merging galaxies and a doubling of the early type fraction going from the outskirts to the intermediate density regions around the cluster.  This suggests that morphological evolution may be especially active in these moderately high-density environments at z = 0.8.  Alternatively, many galaxies in these regions may have already crossed the core and been transformed deep in the cluster.  The high-density cluster core contained nearly 80\% passive early types, lacked spiral galaxies, and was similarly deficient in peculiar galaxies to the intermediate-density regions.  

Galaxies of high stellar mass, on average, had about twice the early-type fraction of galaxies of low stellar mass.  Part of this effect was due to a positive correlation between environmental density and galaxy stellar mass.  To resolve this matter, we looked at at both massive galaxies in the low-density outskirts and the low-mass galaxies in the high- and intermediate-density regions.  The early-type and peculiar fractions in these latter two samples had intermediate values between those typical of high-mass and low-mass samples and high-to-intermediate density and low-density samples.  This result suggests that both galaxy mass and environment (Gobat et al.~\cite{gob08}; Rettura et al.~\cite{ret10}, \cite{ret11}; Muzzin et al.~\cite{muz12}) may affect the morphological evolution of galaxies.

Drastic increases in early-type fraction with environmental density and stellar mass were often paired with decreases in the peculiar, compact, and merging fraction.  This finding suggests that a dense environment may provoke the direct evolution of mostly low-mass peculiar galaxies into early types {\it{instead of}} spirals, as well as turn spirals into early types.  Further detailed morphological studies of clusters in the 0.8 $<$ z $<$ 2 range, integral field spectroscopy, and modeling can further constrain the effects of cluster environments on galaxy evolution.  Perhaps the evolutionary path of infalling cluster galaxies depends on the redshift at which the galaxies fall into the cluster.  Those falling into the cluster at low redshifts or more recent epochs may have the time to first evolve into spirals before evolving into early types.  At higher redshifts, on the other hand, the peculiar galaxies may never get the chance to develop into regular spirals before falling into the cluster.  In the light of our results and the Jaff\'e et al.~(\cite{jaf11}) morphologically regular but kinematically peculiar galaxies falling into clusters, it will be important to explore differences in how gas and stars are affected by environment.  For instance, cold gas stripped from everywhere but the nucleus of an infalling cluster galaxy might create otherwise regular galaxies with blue nuclei.  These galaxies would then become typical early types or spirals again once this final burst of nuclear star formation reaches an age greater than 1 Gyr.

\begin{acknowledgements}
J. N. acknowledges the support provided by FONDECYT postdoctoral research grant N.~3120233.  R.D. gratefully acknowledges the support provided by
the BASAL Center for Astrophysics and Associated Technologies (CATA), and by FONDECYT grant N.~1100540.
This research is based on observations with the Hubble Space Telescope obtained at the Space Telescope Science Institute, operated by the Association of Universities for Research in Astronomy, Inc., under NASA contract NAS 5-26555.  These observations are associated with Program GO-9290.  Spectroscopic data for cluster member confirmation were obtained from FORS2 observations at the ESO VLT as part of programs 166.A-0701, 69.A-0683, 72.A-0759 and 076.A-0889.  We would like to thank M. Postman for providing the full set of visual T-type classifications and a clarification of the definition of early types via private communication.
\end{acknowledgements}


\end{document}